\documentclass[review]{elsarticle}

\makeatletter
\def\ps@pprintTitle{%
	\let\@oddhead\@empty
	\let\@evenhead\@empty
	\def\@oddfoot{}%
	\let\@evenfoot\@oddfoot}
\makeatother

\usepackage{lineno,hyperref}
\usepackage{float}
\modulolinenumbers[20]
\usepackage{amssymb,amsmath,amsthm}
\usepackage{bm}
\usepackage{subcaption}
\usepackage{diagbox}
\usepackage{threeparttable}
\usepackage{color}
\usepackage{natbib}
\usepackage{sectsty}
\usepackage{multirow}
\usepackage{epsfig}
\usepackage[title]{appendix}
\usepackage{tikz}
\usetikzlibrary{intersections}
\usepackage{tcolorbox}
\usepackage{caption}
\usepackage{bm}
\usepackage{rotating}
\usepackage{hyperref}
\usepackage{algorithm}
\usepackage{algpseudocode}

\numberwithin{table}{section}
\numberwithin{equation}{section}
\numberwithin{figure}{section}

\subsectionfont{\small\selectfont}
\sectionfont{\normalsize \selectfont}
\subsubsectionfont{\footnotesize \selectfont}

\newtheorem{prop}{Proposition}[section]
\newtheorem{theorem}{Theorem}[section]

\newtheorem{corollary}{Corollary}[section]
\newtheorem{lemma}[theorem]{Lemma}

\biboptions{authoryear}
\begin{document}
\begin{frontmatter}
\title{When is truncated stop loss optimal?}

\author[address1]{Erik B\o lviken}\ead{erikb@math.uio.no}
\author[address2]{Yinzhi Wang\corref{cor1}}
\ead{yzwang@swufe.edu.cn}
\cortext[cor1]{Corresponding author}

\address[address1]{Department of Mathematics, University of Oslo, Postboks 1053 Blindern, 0316 Oslo,
	Norway}
\address[address2]{Southwestern University of Finance and Economics, 555 Liutai avenue, Chengdu, China}
\begin{abstract}
	The paper examines how reinsurance can be used to strike a
	balance between expected profit and VaR/CVaR risk. Conditions making
	truncated stop loss contracts optimal are derived, and it  is argued that
	those are usually satisfied in practice. One of the prerequisites  is that
	reinsurance is
	not too cheap,
	and an argument resembling arbitrage suggests that it isn't.
	
\end{abstract}

\end{frontmatter}

\section{Introduction}\label{sec:intro}
Reinsurance plays a
vital role in managing risks and improving solvency for
insurance companies and a  wide variety of such contracts have
been introduced, for example quota share, surplus and stop loss.
We are here concerned with the latter type which goes back to the
the pioneering work in \cite{borch1960} with extensions in
\cite{arrow1963,  gajek2000, kaluszka2001,cai2007optimal} and \cite{Hu2015}.
Stop loss in the original Borch form with the reinsurer taking on unlimited
responsibility is rarely used. In practice typically  only the truncated
version is offered, a layer treaty with the reinsurer
carrying all risk in some interval and the insurer the rest.
Whether this is the best way for the insurer to
cede risk which we argue in this paper,
must depend on
the point of view, but suppose
an insurer is using  reinsurance 
to balance expected profit and risk.
This situation was addressed in \cite{chi2017optimal}
through an ingenious mathematical argument utilized in the present paper too.
It was established that 
multi-layer contacts are the best ones
for risk quantified by
Value at Risk (VaR) or Conditional Value at Risk (CVaR). Prerequisites
were Lagrange
(or Markowitz) types of criteria with
reinsurance pricing following the 
market factor model in
\cite{buhlmann1980economic} or, equivalently,
as pointed out in \cite{wang2022much}, the
distortion model in \cite{denneberg1990}. These are
proxies for real reinsurance prices, but many other so-called premium principles
have been proposed, and there is a
long list of them in \cite{young2006premium}; for their applications
in optimal reinsurance,  
consult \cite{kaluszka2001, chi2013optimal, cong2016} and \cite{MENG2023}.
\\\\
We shall here work with the market factor/distortion model.
In its general form this provides a highly flexible
description of reinsurance cost, but not all shapes of the model
are necessarily equally plausible.
Restrictions
will be introduced in Section \ref{sec:2}, and
it will be
argued that they capture
most cases of practical interest. The
optimal multi-layer solutions 
in \cite{chi2017optimal} now reduce to truncated stop loss contracts.
A key part of the argument is the
introduction of
a parameter $\gamma_r$ permitting reinsurance cost to be scaled up and down
in a simple manner and which can be compared to the loading $\gamma$
in the primary market of the cedent. By varying $\gamma_r$ and $\gamma$
over a certain feasible region it will become possible to prove the 
lemma in Section \ref{sec:3} which is  crucial for the insight into
when truncated stop loss contracts are optimal.
\\\\
The Lagrangian approach leads to general multi-layer schemes through a technique is utilized in the present paper. There are in actuarial literature
many other ways to approach optimal reinsurance than Lagrangian approach in \cite{cai2008optimal}. One possibility, 
used
in \cite{arrow1963} and \cite{GUERRA2008}, is to seek the maximum of
expected utility of the net wealth of the cedent. Another general solution
was obtained
by \cite{cui2013optimal} who combined utility with VaR and CVaR risk
and with reinsurance cost described by the the distortion model. This
leads to piecewise-linear reinsurance functions as optima, found also
in \cite{zheng2014optimal,chi2017optimal}
and \cite{lo2017unifying}.
Two other approaches have drawn a lot of
attention  in the literature. One is  the marginal indemnity function (MIF) formulation applied by \cite{Assa2015} and \cite{zhuang2016marginal} and the other
is the empirical approach proposed by
\cite{tan2014empirical} and \cite{sun2017optimal}. The MIF function
is the marginal rate of changes in the value of a reinsurance contract.
As shown by \cite{zhuang2016marginal},  with the MIF formulation,
there is no need to specify a conjecture on the shape
of optimal reinsurance in solving the problem. As uncertainty of
the underlying  distribution has been realized more by researchers,
the empirical approach becomes appealing as a remedy.   The
availability of data and subsequent statistical analysis play an
important role in  formulating the optimal reinsurance model
in empirical approach. \cite{tan2014empirical} proposed a general
framework to formulate a typical optimal reinsurance model and the
solution has been demonstrated robust under some risk
measures. \cite{wang2022much} on the other hand, discuss how
much the optimum procedure is degraded when a model that
is not true is being used. Model uncertainty in optimal reinsurance
design is also examined by \cite{Asimit2017} and \cite{Hu2015}.

\section{Preliminaries}\label{sec:2}
\subsection{Problem and notation}\label{sec:2.1}
Let $X$ be the total claim against a portfolio with $R_I(X)=X-I(X)$ the net risk after   subtracting  $I=I(X)$ in compensation from a reinsurer. Natural restrictions on $I$ are
	\begin{equation}
		0\leq I(x)\leq x
		\hspace*{1cm}\mbox{and}\hspace*{1cm}
		0\leq I(x_2)-I(x_1)\leq x_2-x_1
		\hspace*{0.2cm}\mbox{if}\hspace*{0.2cm}x_1\leq x_2.
		\label{e21}
	\end{equation}
with the second condition known as the incentive compatible condition, which prevents so-called moral hazard; see \cite{Huberman1983,Xu2019,Cai2020}, and \cite{TAN2020}. Premium and reinsurance premium are
\begin{equation}
		\pi=(1+\gamma)E(X)
		\hspace*{1cm}\mbox{and}\hspace*{1cm}
		\pi(I)=E\{I(X)M(Z)\}
		\label{e22}
\end{equation}
	with $\gamma>0$ on the left a fixed loading. This is the well-known expected premium principle. The price regime
	for reinsurance on the right goes back to \cite{buhlmann1980economic}
	and is based on the market factor $M(Z)$ where
	$M(z)$ is a positive function and $Z$ is a positive random variable
	with $E\{M(Z)\}<\infty$.
	\\\\
	Measures of risk are based upon worst case scenarios, namely, Value at Risk (VaR) and Conditional Value at risk (CVaR) in this paper. 
	With $F_{R_I}(x)$ the distribution function of $R_I$, their formal definitions
	at level $\epsilon$ are
	\begin{equation}
		\mbox{VaR}(R_I)=\inf\{x|1-F_{R_I}(x)\leq \epsilon\}
		\quad \mbox{and} \quad
		\mbox{CVaR}(R_I)=E\{R_I(X)|R_I(X)\geq\mbox{VaR}(R_I)\}.
		\label{e23}
	\end{equation} 
They are to be balanced against the expected profit of the insurer which is
	\begin{equation}
		{\cal G}(I)=\gamma E(X)-\{\pi(I)-E\{I(X)\}\}-\beta \rho(R_I)
		\label{e24}
	\end{equation}
	with
	$\rho(R_I)$ a common symbol for VaR and CVaR
	and $\beta\geq 0$ a cost of capital 
	coefficient.  The criterion addressed is the ratio
	\begin{equation}
		{\cal C}(I)=\frac{{\cal G}(I)}{\rho(R_I)},
		\label{e25}
	\end{equation}
	which is to be maximized with respect to all $I$ satisfying~(\ref{e21}). 
	Since ${\cal C}(I)={\cal C}_0(I)-\beta$, where ${\cal C}_0(I)$ is the criterion when $\beta=0$, the optimum $I$ doesn't depend on $\beta$ and $\beta=0$ from now on.
	\\\\
	Special role in this paper is played by the truncated stop loss contract of the form
	\begin{equation}
		I_{ab}(x)=\begin{array}{cl}
			0,&x<a\\
			x-a,&a\leq x\leq b\\
			b-a,&x>b,
		\end{array}
		\label{e31}
	\end{equation}
	for cut-off points $0 < a < b$, with $b=\infty$ resulting in the special case of stop-loss. 
	\subsection{The cost of reinsurance}\label{sec:2.2}
	The risk $X$ and the 
	market factor $M(Z)$ in~(\ref{e22}) right
	can be modeled using a pair of dependent
	uniforms  $(U,V)$. Then
	$X=F^{-1}(U)$ and $Z=G^{-1}(V)$
	where $F^{-1}(u)$ and
	$G^{-1}(v)$ are the inverses of the distribution functions
	of 
	$X$ and $Z$. It 
	is well known that this copula approach
	entails no loss of generality at all. Now
	the reinsurer's expected surplus becomes 
	\begin{equation}
		\pi(I)-E\{I(X)\}=\int_0^\infty K\{F(x)\}dI(x).
		\label{e27}
	\end{equation}
	where
	\begin{equation}
		K(u)=\int_u^1\{E\{M(Z|v)\}-1\}dv.
		\label{e28}
	\end{equation}
Here $K(u)$ is a kernel which
doesn't depend on the 
distribution function
$F(x)$ since the 
conditional mean $E\{M(Z)|v\}$ 
doesn't. Also note that  $K(u)$ is bounded over the interval $[0,1]$
since
\begin{displaymath}
	\int_u^1E\{M(Z|v)\}dv\leq\int_0^1E\{M(Z|v)\}dv=E\{M(Z)\}<\infty.
\end{displaymath}  
	\subsection{An invariance property}\label{sec:2.3}
	The optimum problem is scale invariant.
	To see this introduce for a moment
	notations such as
	$F_\xi(x)$, ${\cal C}_\xi(I)$ and $I_\xi(X)$ rather than $F(x)$,
	${\cal C}(I)$ and $I(X)$ to emphasize  the dependence  on $\xi=E(X)$. Then
	$F_\xi(x)=F_1(x/\xi)$ where $F_1(x)$ has expectation one, and
	any $I_\xi(X)$ under $\xi$ can be linked to
	an equivalent contract $I_1(X)$ under $\xi=1$ through
	\begin{displaymath}
		I_\xi(X)=\xi I_1(X/\xi).  
	\end{displaymath}
	Here either of $I_\xi$ and $I_1$ satisfies~(\ref{e21}) if the other does,
	and $I_\xi$ is simply a rescaled version of $I_1$. For example, when
	$I_1$ is the truncated stop loss contract in~(\ref{e31}) \textbf{above}, $I_\xi$
	is truncated stop loss too, but now with 
	the cut-off points rescaled by $\xi$.
	It is verified in \ref{sec:a} that
	${\cal C}_\xi(I_\xi)={\cal C}_1(I_1)$ which applies
	whether risk is measured by VaR
	or CVaR. This means that all
	results on optimization 
	can be inferred from the case $\xi=1$ which will be utilized in Section \ref{sec:4}.
\subsection{More on reinsurance pricing} \label{sec:2.4}
	The results
	in the next section require additional restrictions on
	reinsurance pricing. The first is to introduce monotone market factor models. Now the conditional expectation $E\{M(Z)|u\}$ increases with $u$ which means
	that high values
	of $u$ associated with 
	high losses $x$ 
	make the perception of future risk higher with
	higher reinsurance cost.
	This condition was proposed
	in
	\cite{wang2022much} who pointed out that
	the equivalent dispersion
	pricing regime is based on a concave distortion function, a common
	assumption; see, for example,
	\cite{cui2013optimal, Assa2015,zhuang2016marginal}
	and \cite{cheung2017characterizations}. This makes the kernel
	$K(u)$ in~(\ref{e28}) and also the normalized  version
	$K_0(u)$ in~(\ref{e210}) below concave and positive curves.
	\\\\
	A second restriction expresses that reinsurance
	isn't too cheap. 
	The function $M(z)$ in the market factor model may without loss
	of generality be specified as
	\begin{equation}
		M(z)=(1+\gamma_r)M_0(z)
		\label{e29}
	\end{equation}
	with $\gamma_r\geq 0$ a parameter and 
	$M_0(z)$ a function for which $E\{M_0(Z)\}=1$. 
	Similar  to~(\ref{e28}) there is the kernel
	\begin{equation}
		K_0(u)=\int_u^1\{E\{M_0(Z)|v\}-1\}dv,
		\label{e210}
	\end{equation}
	and the normalization  $E\{M_0(Z)\}=1$ implies
	$K_0(0)=0$ and for the derivative at the origin
	$0<K'(0)\leq 1$; see Figure \ref{fig1}
	where the function is sketched. The upper bound on $K'(0)$
	is a consequence of  $K_0'(0)=1-\lim_{u\rightarrow 0}E\{M_0(Z)|u\}$ where
	the limit must be less than $1$  since otherwise the monotonicity
	of $E\{M_0(Z)|u\}$ would yield $E\{M_0(Z)\}>1$. 
	\\\\
	It follows from~(\ref{e28}) and~(\ref{e29}) that $\gamma_r=K(0)$. 
	The parameter is a 
	`loading' in the reinsurance market
	when the reinsurer contributes $I(x)=1$
	regardless what the loss is. Much 
	literature in optimal reinsurance assumes $\gamma_r=0$ which goes
	back to \cite{buhlmann1980economic} who offered an economic
	equilibrium argument in support,
	but there is also a different viewpoint.
	Consider
	a portfolio reinsured so heavily
	that the net reserve of the insurer is zero; the reinsurance
	contract of such an arrangement will be denoted $I_{0x_\epsilon}$ below. 
	It is for a portfolio of independent risks  verified in \ref{sec:b} that
	\begin{equation}
		\frac{{\cal G}(I_{0x_\epsilon})}{\xi}
		\longrightarrow \gamma-\gamma_r
		\hspace*{1cm}\mbox{as}\hspace*{1cm}\xi\rightarrow\infty,
		\label{e212}
	\end{equation}
	where $\xi=E(X)$.
	Hence,
	unless $\gamma_r\geq \gamma$, the insurer
	would for large $\xi$ obtain huge profit with zero reserve.
	This is not quite arbitrage
	since the risk above $x_\epsilon$ is still there, but is it likely that
	insurance and reinsurance
	markets  allow such a state of affairs? We think not,
	and shall assume 
	$\gamma_r\geq \gamma$. It will also be taken for granted that
	\begin{equation}
		{\cal G}(I_{0x_\epsilon})\leq 0
		\label{e213}
	\end{equation}
	for any $\xi$. If this isn't true,
	the maximization of~(\ref{e25}) is trivial
	since ${\cal C}(I_{0x_\epsilon})$ is infinite.
	
\section{Results}\label{sec:3}
For the truncated stop loss contract $I_{ab}$ introduced in Section \ref{sec:2}
 there are the following results proved in Sections \ref{sec:4.1} and \ref{sec:4.2}:
  \begin{prop}\label{prop1}
  Suppose $E\{M(Z|u\}$ is an upwards
  monotone function of $u$. 
  Then a stop loss contract maximizes
  ${\cal C}(I)$ among all $I$
  satisfying~(\ref{e21}) if and only if
  $\sup_{a,b}{\cal C}(I_{ab})\leq \gamma_r$,
  and this  applies
  whether ${\cal C}(I)$ is based on VaR or CVaR.
  \end{prop}
  \noindent
  That $x_\epsilon\geq E(X)$ is needed for the
  proof of \textit{the only if part} in Section \ref{sec:4}.
  We
  must study when $\gamma_r$ is an upper bound on all ${\cal C}(I_{ab})$.
  The following lemma is proved in Section \ref{sec:4.3}.
\begin{lemma}\label{lemma1}
Suppose  $E\{M_0(Z|u\}$ is an upwards monotone function of $u$,
$\gamma_r\geq \gamma$, $x_\epsilon\geq E(X)$  and
\begin{equation}
K_0'(0)\int_{x_\epsilon}^\infty\{1-F(x)\}dx\leq \int_0^{x_{\epsilon}}K_0\{F(x)\}dx.
\label{e33}
\end{equation}
Then   $\sup_{a,b}{\cal C}(I_{ab})\leq \gamma_r$
for both the VaR and CVaR ratios.
\end{lemma} 
\noindent
Together Proposition \ref{prop1} and Lemma \ref{lemma1} yield the following corollary:
\begin{corollary}\label{coro1}
Under the conditions of Lemma \ref{lemma1}
a stop loss contract maximizes
${\cal C}(I)$ among all $I$
satisfying~(\ref{e21}). This
applies  whether ${\cal C}(I)$ is based on VaR or CVaR.
\end{corollary}
\noindent
Truncated stop loss contracts are not optimal if the technical
condition~(\ref{e33}) does not hold which emerges from the proof in Section \ref{sec:4.3}. 
	The condition $x_\epsilon\geq  E(X)$ is unproblematic, and we
	conjecture that~(\ref{e33}) usually holds too.
	To get a feeling for it
	imitate a real situation by assuming
	$K_0(u)=c(u-u^2)$ where $c$ is a coefficient with $0<c\leq 1$. Then 
	$K_0(0)=K_0(1)=0$, $K'(0)=c$  with $K_0(u)$ positive and concave, 
	the
	properties of real kernels.
	If $F(x)=1-e^{-x}$
	is the distribution function of the exponential
	distribution, then elementary calculations show that
	the left hand side of~(\ref{e33}) is $c\epsilon$ and the right hand side
	$c(1-\epsilon)^2/2$,
	convincingly much higher since $\epsilon$ is a small number.
	Other, less heavy-tailed  distributions
	lead to the same conclusion when calculated numerically.
\section{Conclusion}\label{sec:5}
	When are truncated stop loss contracts optimal? In one word
	almost always when the aim is to balance profit against VaR or CVaR risk.
	A prerequisite is that reinsurance is priced through a
	monotone market factor model. 
	The conclusion fails when reinsurance is priced too low,
	but the arbitrage-like argument in Section \ref{sec:2.2} suggests that this may not be likely in practice.
\section{Proofs}\label{sec:4}
 \subsection{Proposition \ref{prop1} for VaR risk}\label{sec:4.1}
 {\em The Lagrange optimum.} Consider the
 Lagrangian set-up
	\begin{equation}
		{\cal L}(I)={\cal G}(I)-\mu\rho(I)
		\label{e51}
	\end{equation}
	with $\mu$ a coefficient. Using calculations in
	\cite{chi2017optimal} (added for completeness in \ref{sec:b})
	${\cal L}(I)$ can be rewritten as
	\begin{equation}
		{\cal L}(I)=\gamma E(X)-\mu x_\epsilon +\int_0^\infty\psi(x)dI(x),
		\label{e52}
	\end{equation}
	where
	\begin{equation}
		\psi(x)=\mu{\cal I}_{x<x_\epsilon}-K\{F(x)\}.
		\label{e53}
	\end{equation}
	Here  ${\cal L}(I)$ depends on $I$ through the 
	integral on the very right and is maximized 
	when $I(x)=1$ for
	$\psi(x)\geq 0$ and
	$I(x)=0$ otherwise since~(\ref{e21}) right
	implies $0\leq dI(x)/dx\leq 1$ everywhere where the derivative exists.
	This result, due to \cite{chi2017optimal}, shows that the
	Lagrange optimum $I$
	is a multi-layer contract defined by the fluctuations
	of $\psi(x)$ around $0$.
	\\\\
 {\em The if part.} Let $I_{ax_\epsilon}$ be the stop loss
reinsurance function with
$x_\epsilon$ as the upper cut-off point and
define $\hat{a}\leq x_\epsilon$ as a solution of the equation
\begin{equation}
	K\{F(\hat{a})\}={\cal C}(I_{\hat{a}x_\epsilon});
	\label{e54}
\end{equation}
if no such $\hat{a}$ exists let $\hat{a}=x_\epsilon$. Suppose
$\mu={\cal C}(I_{\hat{a}x_\epsilon})$ in~(\ref{e53}).
Then by  assumption
\begin{displaymath}
	\psi(0)=\mu-K(0)={\cal C}(I_{\hat{a}x_\epsilon})-\gamma_r\leq 0
	\hspace*{0.6cm}\mbox{and also}\hspace*{0.6cm}
	\psi(\hat{a})={\cal C}(I_{\hat{a}x_\epsilon})-K\{F(\hat{a})\}= 0. 
\end{displaymath}
It follows that
$\psi(x)$ starts out $ \leq 0$, then becomes $\geq 0$
from $x=\hat{a}$ until $x=x_\epsilon$ before returning to $\leq 0$
above $x_\epsilon$. This is because $K(u)$ in~(\ref{e53}) is 
either decreasing everywhere or first increasing and then decreasing
and never negative.
But
the characterization of the Lagrange optimum above
now implies that
$I_{\hat{a}x_\epsilon}$ maximizes ${\cal L}(I)$
among all $I$ satisfying~(\ref{e21}). Hence
\begin{displaymath}
	{\cal G}(I)-\mu\rho(I)\leq
	{\cal G}(I_{\hat{a}x_\epsilon})-\mu\rho(I_{\hat{a}x_\epsilon}),
\end{displaymath}
or when $\mu={\cal C}(I_{\hat{a}x_\epsilon})$ is inserted
\begin{displaymath}
	{\cal G}(I)-{\cal C}(I_{\hat{a}x_\epsilon})\rho(I)\leq 0
	\hspace*{0.6cm}\mbox{which yields}\hspace*{0.6cm}
	\{{\cal C}(I)-{\cal C}(I_{\hat{a}x_\epsilon})\}\rho(I)\leq 0.
\end{displaymath}
Since $\rho(I)>0$, it follows that
${\cal C}(I)\leq {\cal C}(I_{\hat{a}x_\epsilon})$ which was to be proved.
	\\\\
	{\em An auxiliary result.} The proof of the only if part
	makes use of 
	\begin{equation}
		{\cal C}(I_{0b})\leq\gamma_r
		\hspace*{1cm}\mbox{when}\hspace*{1cm}0\leq b\leq x_\epsilon.
		\label{e54a}
	\end{equation}  
	To verify this
	note that ${\cal G}(I_{0b})=\gamma\xi-\int_0^bK\{F(x)\}dx$
	and $\rho(I_{0b})=x_\epsilon-b$ with ${\cal C}(I_{0b})$ their ratio. The inequality
	is thus equivalent to
	\begin{displaymath}
		-\gamma_r b+\int_0^b K\{F(x)\}dx\geq \gamma\xi-\gamma_rx_\epsilon
	\end{displaymath}
	which is true when $b=0$ since it has been assumed that
	$x_\epsilon\geq \xi$ and $\gamma_r\geq \gamma$ and when $b=x_\epsilon$
	because of~(\ref{e213}). It is also true for all $b$ in between
	since the
	derivative of the left hand side is
	$-\gamma_r+K\{F(b)\}=-K(0)+K\{F(b)\}$ which is $\geq 0$  for small $b$
	(if at all) and $\leq 0$ afterwards.
	\\\\
	{\em The only if part.} Suppose ${\cal C}(I_{\hat{a}\hat{b}})>\gamma_r$ where
	$I_{\hat{a}\hat{b}}$ maximizes
	${\cal C}(I_{ab})$ among all $I_{ab}$.
	Let
	$\mu={\cal C}(I_{\hat{a}\hat{b}})$ and note that 
	$\psi(0)>0$ in~(\ref{e53}) which implies that
	the reinsurance function $\hat{I}$
	maximizing ${\cal L}(I)$
	for this value of $\mu$
	must have a layer starting at the origin.
	Now
	\begin{displaymath}
		0={\cal G}(I_{\hat{a}\hat{b}})-{\cal C}(I_{\hat{a}\hat{b}})\rho(I_{\hat{a}\hat{b}})\leq
		{\cal G}(\hat{I})-{\cal C}(I_{\hat{a}\hat{b}})\rho(\hat{I})
		\hspace*{1cm}\mbox{or}\hspace*{1cm}                             
		\{{\cal C}(\hat{I})-{\cal C}(I_{\hat{a}\hat{b}})\}\rho(\hat{I})\geq 0,
	\end{displaymath}
	so that ${\cal C}(\hat{I})\geq{\cal C}(I_{\hat{a}\hat{b}})$.
	But then
	${\cal C}(\hat{I})>\gamma_r\geq {\cal C}(I_{0b})$ by~(\ref{e54a}),
	and $\hat{I}$ differs from $I_{0b}$ for all $b$
	which means that $\hat{I}$ has at least
	one more layer.
\subsection{Proposition \ref{prop1}  for CVaR}\label{sec:4.2}
	The proof for the CVaR part of the proposition  follows the argument in
	Section \ref{sec:4.1} except for some details. The representation of 
	${\cal L}(I)$ in~(\ref{e52}) and~(\ref{e53}) now changes to
	\begin{equation}
		{\cal L}(I)=\gamma E(X)-\mu E(X|X\geq x_\epsilon)+\int_0^\infty \psi(x)dI(x)
		\label{e55}  
	\end{equation}
	where
	\begin{equation}
		\psi(x)=\mu{\cal I}_{x\leq x}-K\{F(x)\}+
		\frac{\mu}{\epsilon}\{1-F(x)\}{\cal I}_{x\geq x},
		\label{e56}  
	\end{equation}
	which is verified in \ref{sec:b}. Note that~(\ref{e56}) coincides 
	with~(\ref{e53}) below $x_\epsilon$ whereas the additional term
	on the very right appears when $x>x_\epsilon$.
	It follows that \textit{the if part} is the same as in Section \ref{sec:4.1} except
	for $\psi(x)$ being positive above $x_\epsilon$, possibly up to infinity.
	\textit{The only if part} is similar to the proof in Section \ref{sec:4.1}. In
	particular~(\ref{e54a}) is still true on the
	same argument as above with
	$\rho(I_{0b})=E(X|X\geq x_\epsilon)-b$ now replacing
	$\rho(I_{0b})=x_\epsilon-b$.
\subsection{Lemma \ref{lemma1}}\label{sec:4.3}
\begin{figure}[H]
		\centering
		\begin{tikzpicture}[scale=1.2]
			\draw[->](0,0) -- (9,0) node[below] {$u$};
			\draw[->](0,0) -- (0,5) node[left] { $K_0(u)$};
			\node[below left]at(0,0) {\footnotesize $0$};
			\draw[black,line width=1pt,name path=plot](0,0)..controls(1,0.9) and(2,1.6)..(3,1.8)..	controls (4,2) and (5,1.8)..(6,1.5).. controls(7,1.1) and (7.5,0.8)..(8.2,0);
			\node[below](a)at(4,0){$u_\gamma=F(a_\gamma)$};
			\draw[dashed] (3.8,0)--(3.8,1.9) ;
			\node[below](a)at(6,0){
				$u_{\underline{\gamma}}=F(1-\epsilon)$};
			\draw[dashed] (6,0)--(6,1.5) ;
			\draw[dashed] (8.2,0)--(8.2,0) node[below]{1};
			\path[draw, name path= upward line] (0,0) -- (7,3.45) node[right] {\large$\frac{\gamma}{1+\gamma}u$};
			\path[draw, name path= upward line] (0,0) -- (8,2)node[right] {\large$\frac{\underline{\gamma}}{1+\underline{\gamma}}u$};
			\path[draw, name path= upward line] (0,0) -- (5,4) node[right] {\large$\frac{\overline{\gamma}}{1+\overline{\gamma}}u$};
		\end{tikzpicture}
		\caption{Plot of $K_0(u)$ with $a_\gamma$,
			$\underline{\gamma}$ and $\overline{\gamma}$ as defined
			in the text.}
			\label{fig1}
	\end{figure}
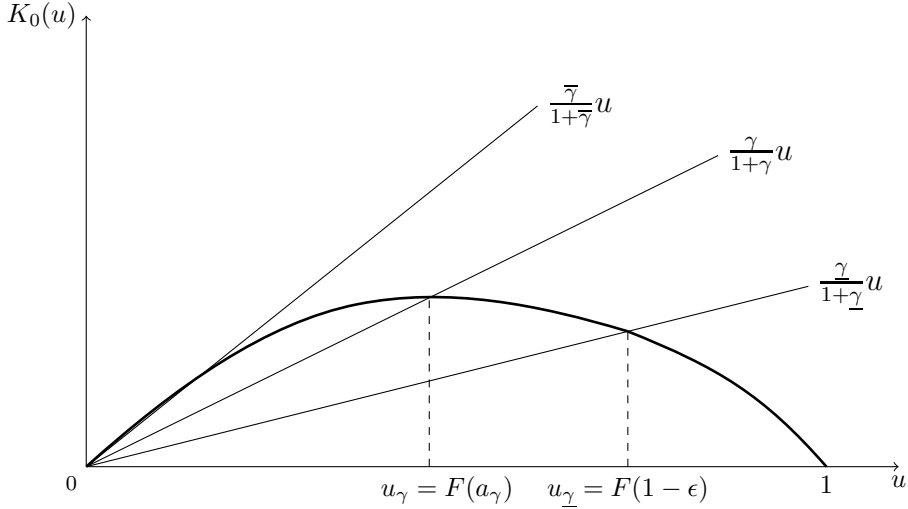

	{\em Preliminaries.} The optimum among all stop loss contracts is
	$I_{\hat{a}x_\epsilon}$ where $\hat{a}$ satisfies~(\ref{e54}), and the
	assertion of Lemma \ref{lemma1} is 
	${\cal C}(I_{\hat{a}x_\epsilon})\leq \gamma_r$.
	Only
	VaR needs to be considered since $\mbox{CVaR}\geq\mbox{VaR}$ so that
	${\cal C}(I)$
	is always smaller for CVaR. We may also assume
	$E(X)=1$ since
	the criterion does not depend on $\xi=E(X)$. 
	The two kernels $K(u)$ and $K_0(u)$ in Section \ref{sec:2} have a simple relationship
	which will be used below.
	Insert~(\ref{e29}) into~(\ref{e28}) and utilize~(\ref{e210}).  This yields
	\begin{equation}
		K(u)=(1+\gamma_r)K_0(u)+\gamma_r(1-u),
		\label{e57}
	\end{equation}
	and since both terms on the right are positive,
	$K(u)$ increases with $\gamma_r$ and 
	reinsurance cost too so that
	${\cal G}(I)$ and hence ${\cal C}(I)$ go down for all $I$.
	But 
	$\gamma_r=\gamma$ is then a worst-case scenario. If it can be proved that 
	${\cal C}(I_{\hat{a}x_\epsilon})\leq \gamma_r$ when $\gamma_r=\gamma$,
	then this must hold for all larger values of $\gamma_r$ as well.
	\\\\
	{\em A key condition.} It may thus be assumed that $\gamma_r=\gamma$.
	This common value will be varied
	which makes it convenient to introduce notations such as
	${\cal C}(I;\gamma)$,
	${\cal G}(I;\gamma)$ and $K(u;\gamma)$
	instead of ${\cal C}(I)$,
	${\cal G}(I)$ and $K(u)$
	and also $\hat{a}_\gamma$ rather than $\hat{a}$
	to highlight the dependence on $\gamma$.
The equation~(\ref{e54}) for $\hat{a}_\gamma$
now reads $K\{F(\hat{a}_\gamma);\gamma\}= {\cal C}(I_{\hat{a}_\gamma x_\epsilon})$,
and
when $\xi=1$ becomes
	\begin{displaymath}
		K\{F(\hat{a}_\gamma);\gamma\}=
		\frac{{\cal G}(I_{\hat{a}_\gamma x_\epsilon};\gamma)}{\hat{a}_\gamma}
		\hspace*{0.7cm}\mbox{where}\hspace*{0.7cm}
		{\cal G}(I_{\hat{a}_\gamma x_\epsilon};\gamma)=
		\gamma-\int_{\hat{a}_\gamma}^{x_\epsilon}K\{F(x);\gamma\}dx.
	\end{displaymath}
	Hence, when introducing
	\begin{equation}
		H(a;\gamma)=a K\{F(a);\gamma\}+\int_{a}^{x_\epsilon}K\{F(x);\gamma\}dx,
		\label{e58}
	\end{equation}
	where $\hat{a}_\gamma$ is the solution of
	$H(\hat{a}_\gamma;\gamma)=\gamma$.
	We are assuming $K(0)=\gamma$, and if $K(u)$ is downwards monotone,
	then~(\ref{e54}) yields 
	${\cal C}(I_{\hat{a}x_\epsilon})\leq \gamma$
	which is the assertion of Lemma \ref{lemma1}. In the opposite case
	where $K(u)$ first goes up from $K(0)=\gamma$ and then
	down to $K(1)=0$ there is some $a_\gamma>0$
	so that 
	\begin{equation}
		K\{F(a_\gamma);\gamma\}=\gamma
		\label{e59}
	\end{equation}
	with derivative $\partial K\{F(a);\gamma\}/\partial a\leq 0$
	and $\partial K\{F(a);\gamma\}/\partial a\leq 0$ when $a\geq a_\gamma$.
	But from~(\ref{e58})
	\begin{displaymath}  
		\frac{\partial H(a;\gamma)}{\partial a}=
		a\frac{\partial K\{F(a);\gamma\}}{\partial a},
	\end{displaymath}
	so that $\partial H(a;\gamma)/\partial a\leq 0$
	when $a\geq a_\gamma$ too.
	Consider the inequality
	\begin{equation}
		H(a_\gamma;\gamma)\geq\gamma,
		\label{e59a}
	\end{equation}
	which plays a key role in the argument. 
	By~(\ref{e213}) $H(0;\gamma)=\int_0^{x_\epsilon}K\{F(x)\}dx\geq \gamma$,
	and if~(\ref{e59a}) is true,
	the
	solution $\hat{a}_\gamma$ of the equation $H(\hat{a}_\gamma;\gamma)=\gamma$
	must be to the right of $a_\gamma$ 
	where $K\{F(a);\gamma\} $ are decreasing in $a$. But then
	\begin{displaymath} 
		{\cal C}(I_{\hat{a}_\gamma x_\epsilon})=K\{F(\hat{a}_\gamma);\gamma\}\leq K\{F(a_\gamma);\gamma\}=\gamma,
	\end{displaymath}
	which is the assertion of Lemma \ref{lemma1}.
	 \\\\
	{\em The inequality~(\ref{e59a}).}
	Insert~(\ref{e57}) with $\gamma_r=\gamma$
	into~(\ref{e59}) 
	which implies that $a_\gamma$ satisfies
	\begin{equation}
		K_0\{F(a_\gamma)\}=\frac{\gamma}{1+\gamma}F(a_\gamma)
		\label{e511}
	\end{equation}
	as depicted in Figure \ref{fig1}. The function $K_0(u)$ starts and ends
	 at zero and its derivative 
	 $K'_0(0)$ at the origin satisfies
	 $0<K'(0)\leq 1$ as remarked in Section \ref{sec:2.4}.
	 But $K_0(u)$ is
	 concave by assumption which implies that~(\ref{e511})
	 has a single solution for all $\gamma\leq\overline{\gamma}$
	 where
	
	\begin{displaymath}
		\frac{\overline{\gamma}}{1+\overline{\gamma}}=K'_0(0).
	\end{displaymath}
	It is clear from Figure \ref{fig1} that $a_{\overline{\gamma}}=0$. At the other end
	the largest $a$ of interest is
	$a_{\underline{\gamma}}=x_\epsilon$ which is the solution of~(\ref{e511}) for
	the smallest value  
	$\underline{\gamma}=K_0(1-\epsilon)$ of $\gamma$,
	and as $\gamma$ varies between
	$\underline{\gamma}$ and $\overline{\gamma}$ the corresponding $a_\gamma$
	covers the interval from $0$ to $x_\epsilon$ exactly.
	\\\\
	We shall first verify~(\ref{e59a}) at the end points
	$\underline{\gamma}$ and $\overline{\gamma}$. 
	Inserting $a_{\underline{\gamma}}=x_\epsilon$
	into~(\ref{e58}) and utilizing
	$K\{F(a_{\underline{\gamma}};\underline{\gamma})\}=\underline{\gamma}$
	yield
	$ H(a_{\underline{\gamma}};\underline{\gamma})
	=x_\epsilon\underline{\gamma}\geq\underline{\gamma}$
	since $x_\epsilon\geq E(X)=1$.
	At the other end point $a_{\overline{\gamma}}=0$ we have
	\begin{displaymath}
		H(a_{\overline{\gamma}};\overline{\gamma})=\int_{0}^{x_\epsilon}K\{F(x);\overline{\gamma}\}dx
		=(1+{\overline{\gamma}})\int_{0}^{x_\epsilon}K_0\{F(x)\}dx
		+{\overline{\gamma}}\int_0^{x_\epsilon}\{1-F(x)\}dx,
	\end{displaymath}
	or since $\int_0^\infty\{1-F(x)\}dx=E(X)=1$, this leads to
	\begin{displaymath}
		H(a_{\overline{\gamma}};\overline{\gamma})
		=(1+{\overline{\gamma}})\int_{0}^{x_\epsilon}K_0\{F(x)\}dx
		-{\overline{\gamma}}\int_{x_\epsilon}^\infty\{1-F(x)\}dx+\overline{\gamma},
	\end{displaymath}
	and now~(\ref{e33}) implies
	$H(a_{\overline{\gamma}};\overline{\gamma})\geq \overline{\gamma}$ 
	since
	$K'(0)=\overline{\gamma}/(1+\overline{\gamma})$.
	
\noindent
The inequality~(\ref{e59a})  and hence the lemma  now follows
if it can be established that $H(a_\gamma;\gamma)$
is concave  
between the two end points of $\gamma$. First note that~(\ref{e57})
with $\gamma_r=\gamma$ yields
\begin{displaymath}
\int_{a}^{x_\epsilon}K\{F(x);\gamma\}dx=(1+\gamma)\int_{a}^{x_\epsilon}K_0\{F(x)\}dx
+\gamma\int_{a}^{x_\epsilon}\{(1-F(x)\}dx,
\end{displaymath}
which by~(\ref{e58}) and~(\ref{e59}) implies
\begin{displaymath}
H(a_\gamma;\gamma)=(1+\gamma)\int_{a_\gamma}^{x_\epsilon}K_0\{F(x)\}dx
+\gamma\int_{a_\gamma}^{x_\epsilon}\{1-F(x)\}dx+\gamma a_\gamma.
\end{displaymath}
Its derivative is
\begin{displaymath}
\frac{dH(a_\gamma;\gamma)}{d\gamma}=\int_{a_\gamma}^{x_\epsilon}K_0\{F(x)\}dx+
\int_{a_\gamma}^{x_\epsilon}\{1-F(x)\}dx+a_\gamma
\end{displaymath}
\vspace*{-0.2cm}
\begin{displaymath}
\hspace*{2.2cm}+\Big(-(1+\gamma)K_0\{F(a_\gamma)\}-\gamma\{1-F(a_\gamma)\}+\gamma\Big)
\frac{d a_\gamma}{d\gamma},
\end{displaymath}
where the second line vanishes because of~(\ref{e511}), and
a new round of differentiation of the first line yields
\begin{displaymath}
\frac{d^2 H(a_\gamma;\gamma)}{d \gamma^2}
=\Big(-K_0\{F(a_\gamma)\}-\{1-F(a_\gamma)\}+1\Big)
\frac{da_\gamma}{d\gamma}.
\end{displaymath}
When again using~(\ref{e511}) to replace $-K_0\{F(a_\gamma)\}$, this implies
\begin{displaymath}
\frac{d^2 H(a_\gamma;\gamma)}{d \gamma^2} 
=\frac{F(a_\gamma)}{1+\gamma}\,\frac{d a_\gamma}{d\gamma}\leq 0,
\end{displaymath}
since the geometry in Figure \ref{fig1}
shows that $da_\gamma/d\gamma\leq 0$. The proof of Lemma \ref{lemma1} is now complete.
\\\\
{\em If condition~(\ref{e33}) does not hold.}    
It emerges from the proof of~(\ref{e59a}) that
$H(a_{\overline{\gamma}};\overline{\gamma})<\overline{\gamma}$ when~(\ref{e33})
is not satisfied. This
makes the situation
the opposite of that above
in that
$\hat{a}_{\overline{\gamma}}< a_{\overline{\gamma}}$ so that
\begin{displaymath}
{\cal C}(I_{\hat{a}_{\overline{\gamma}} x_\epsilon};\overline{\gamma})=
K\{F(\hat{a}_{\overline{\gamma}});\overline{\gamma}\}
>K\{F(a_{\overline{\gamma}};\overline{\gamma}\}=\overline{\gamma}.
\end{displaymath}
Hence Lemma \ref{lemma1} isn't true  at the boundary $\overline{\gamma}$, thus 
nor when $\gamma$ and $\gamma_r$ are close to $\overline{\gamma}$, and when
the lemma isn't satisfied, Proposition \ref{prop1} in Section \ref{sec:3} tells us that
the optimum for these values of
$\gamma$ and $\gamma_r$ must include more than one layer.

\section*{Declaration of competing interest}
The authors declare that there are no competing interests.

\section*{Acknowledgments}
The authors are very grateful to those who provided valuable comments on this article.

\bibliography{optre24.bib}

\begin{thebibliography}{31}
\expandafter\ifx\csname natexlab\endcsname\relax\def\natexlab#1{#1}\fi
\providecommand{\url}[1]{\texttt{#1}}
\providecommand{\href}[2]{#2}
\providecommand{\path}[1]{#1}
\providecommand{\DOIprefix}{doi:}
\providecommand{\ArXivprefix}{arXiv:}
\providecommand{\URLprefix}{URL: }
\providecommand{\Pubmedprefix}{pmid:}
\providecommand{\doi}[1]{\href{http://dx.doi.org/#1}{\path{#1}}}
\providecommand{\Pubmed}[1]{\href{pmid:#1}{\path{#1}}}
\providecommand{\bibinfo}[2]{#2}
\ifx\xfnm\relax \def\xfnm[#1]{\unskip,\space#1}\fi
\bibitem[{Arrow(1963)}]{arrow1963}
\bibinfo{author}{Arrow, K.J.}, \bibinfo{year}{1963}.
\newblock \bibinfo{title}{Uncertainty and the welfare economics of medical
  care}.
\newblock \bibinfo{journal}{The American economic review} \bibinfo{volume}{53},
  \bibinfo{pages}{941--973}.
\bibitem[{Artzner et~al.(1999)Artzner, Delbaen, Eber and Heath}]{Artzner1999}
\bibinfo{author}{Artzner, P.}, \bibinfo{author}{Delbaen, F.},
  \bibinfo{author}{Eber, J.M.}, \bibinfo{author}{Heath, D.},
  \bibinfo{year}{1999}.
\newblock \bibinfo{title}{Coherent measures of risk}.
\newblock \bibinfo{journal}{Mathematical Finance} \bibinfo{volume}{9},
  \bibinfo{pages}{203--228}.
\newblock \DOIprefix\doi{10.1111/1467-9965.00068}.
\bibitem[{Asimit et~al.(2017)Asimit, Bignozzi, Cheung, Hu and Kim}]{Asimit2017}
\bibinfo{author}{Asimit, A.V.}, \bibinfo{author}{Bignozzi, V.},
  \bibinfo{author}{Cheung, K.C.}, \bibinfo{author}{Hu, J.},
  \bibinfo{author}{Kim, E.S.}, \bibinfo{year}{2017}.
\newblock \bibinfo{title}{Robust and pareto optimality of insurance contracts}.
\newblock \bibinfo{journal}{European Journal of Operational Research}
  \bibinfo{volume}{262}, \bibinfo{pages}{720--732}.
\newblock \DOIprefix\doi{10.1016/j.ejor.2017.04.029}.
\bibitem[{Assa({2015})}]{Assa2015}
\bibinfo{author}{Assa, H.}, \bibinfo{year}{{2015}}.
\newblock \bibinfo{title}{{On optimal reinsurance policy with distortion risk
  measures and premiums}}.
\newblock \bibinfo{journal}{Insurance: Mathematics and Economics}
  \bibinfo{volume}{{61}}, \bibinfo{pages}{{70--75}}.
\bibitem[{Boonen and Ghossoub(2021)}]{BOONEN2021}
\bibinfo{author}{Boonen, T.J.}, \bibinfo{author}{Ghossoub, M.},
  \bibinfo{year}{2021}.
\newblock \bibinfo{title}{Optimal reinsurance with multiple reinsurers:
  Distortion risk measures, distortion premium principles, and heterogeneous
  beliefs}.
\newblock \bibinfo{journal}{Insurance: Mathematics and Economics}
  \bibinfo{volume}{101}, \bibinfo{pages}{23--37}.
\newblock \URLprefix
  \url{https://www.sciencedirect.com/science/article/pii/S0167668720300883},
  \DOIprefix\doi{https://doi.org/10.1016/j.insmatheco.2020.06.008}.
  \bibinfo{note}{behavioral Insurance: Mathematics and Economics}.
\bibitem[{Borch(1960)}]{borch1960}
\bibinfo{author}{Borch, K.}, \bibinfo{year}{1960}.
\newblock \bibinfo{title}{An attempt to determine the optimum amount of stop
  loss reinsurance} .
\bibitem[{B{\"u}hlmann(1980)}]{buhlmann1980economic}
\bibinfo{author}{B{\"u}hlmann, H.}, \bibinfo{year}{1980}.
\newblock \bibinfo{title}{An economic premium principle}.
\newblock \bibinfo{journal}{ASTIN Bulletin: The Journal of the IAA}
  \bibinfo{volume}{11}, \bibinfo{pages}{52--60}.
\bibitem[{Cai and Chi(2020)}]{Cai2020}
\bibinfo{author}{Cai, J.}, \bibinfo{author}{Chi, Y.}, \bibinfo{year}{2020}.
\newblock \bibinfo{title}{Optimal reinsurance designs based on risk measures: a
  review}.
\newblock \bibinfo{journal}{Statistical Theory and Related Fields}
  \bibinfo{volume}{4}, \bibinfo{pages}{1--13}.
\newblock \URLprefix \url{https://doi.org/10.1080/24754269.2020.1758500},
  \DOIprefix\doi{10.1080/24754269.2020.1758500},
  \href{http://arxiv.org/abs/https://doi.org/10.1080/24754269.2020.1758500}{{\tt
  arXiv:https://doi.org/10.1080/24754269.2020.1758500}}.
\bibitem[{Cai and Tan(2007)}]{cai2007optimal}
\bibinfo{author}{Cai, J.}, \bibinfo{author}{Tan, K.S.}, \bibinfo{year}{2007}.
\newblock \bibinfo{title}{Optimal retention for a stop-loss reinsurance under
  the var and cte risk measures}.
\newblock \bibinfo{journal}{ASTIN Bulletin: The Journal of the IAA}
  \bibinfo{volume}{37}, \bibinfo{pages}{93--112}.
\bibitem[{Cai et~al.(2008)Cai, Tan, Weng and Zhang}]{cai2008optimal}
\bibinfo{author}{Cai, J.}, \bibinfo{author}{Tan, K.S.}, \bibinfo{author}{Weng,
  C.}, \bibinfo{author}{Zhang, Y.}, \bibinfo{year}{2008}.
\newblock \bibinfo{title}{Optimal reinsurance under var and cte risk measures}.
\newblock \bibinfo{journal}{Insurance: Mathematics and Economics}
  \bibinfo{volume}{43}, \bibinfo{pages}{185--196}.
\bibitem[{Cheung and Lo(2017)}]{cheung2017characterizations}
\bibinfo{author}{Cheung, K.C.}, \bibinfo{author}{Lo, A.}, \bibinfo{year}{2017}.
\newblock \bibinfo{title}{Characterizations of optimal reinsurance treaties: a
  cost-benefit approach}.
\newblock \bibinfo{journal}{Scandinavian Actuarial Journal}
  \bibinfo{volume}{2017}, \bibinfo{pages}{1--28}.
\bibitem[{Chi et~al.(2017)Chi, Lin and Tan}]{chi2017optimal}
\bibinfo{author}{Chi, Y.}, \bibinfo{author}{Lin, X.S.}, \bibinfo{author}{Tan,
  K.S.}, \bibinfo{year}{2017}.
\newblock \bibinfo{title}{Optimal reinsurance under the risk-adjusted value of
  an insurer’s liability and an economic reinsurance premium principle}.
\newblock \bibinfo{journal}{North American Actuarial Journal}
  \bibinfo{volume}{21}, \bibinfo{pages}{417--432}.
\bibitem[{Chi and Tan(2013)}]{chi2013optimal}
\bibinfo{author}{Chi, Y.}, \bibinfo{author}{Tan, K.S.}, \bibinfo{year}{2013}.
\newblock \bibinfo{title}{Optimal reinsurance with general premium principles}.
\newblock \bibinfo{journal}{Insurance: Mathematics and Economics}
  \bibinfo{volume}{52}, \bibinfo{pages}{180--189}.
\bibitem[{Cong and Tan(2016)}]{cong2016}
\bibinfo{author}{Cong, J.}, \bibinfo{author}{Tan, K.S.}, \bibinfo{year}{2016}.
\newblock \bibinfo{title}{Optimal var-based risk management with reinsurance}.
\newblock \bibinfo{journal}{Annals of Operations Research}
  \bibinfo{volume}{237}, \bibinfo{pages}{177--202}.
\bibitem[{Cui et~al.(2013)Cui, Yang and Wu}]{cui2013optimal}
\bibinfo{author}{Cui, W.}, \bibinfo{author}{Yang, J.}, \bibinfo{author}{Wu,
  L.}, \bibinfo{year}{2013}.
\newblock \bibinfo{title}{Optimal reinsurance minimizing the distortion risk
  measure under general reinsurance premium principles}.
\newblock \bibinfo{journal}{Insurance: Mathematics and Economics}
  \bibinfo{volume}{53}, \bibinfo{pages}{74--85}.
\bibitem[{Denneberg(1990)}]{denneberg1990}
\bibinfo{author}{Denneberg, D.}, \bibinfo{year}{1990}.
\newblock \bibinfo{title}{Premium calculation: why standard deviation should be
  replaced by absolute deviation}.
\newblock \bibinfo{journal}{ASTIN Bulletin: The Journal of the IAA}
  \bibinfo{volume}{20}, \bibinfo{pages}{181--190}.
\bibitem[{Gajek and Zagrodny(2000)}]{gajek2000}
\bibinfo{author}{Gajek, L.}, \bibinfo{author}{Zagrodny, D.},
  \bibinfo{year}{2000}.
\newblock \bibinfo{title}{Insurer’s optimal reinsurance strategies}.
\newblock \bibinfo{journal}{Insurance: Mathematics and Economics}
  \bibinfo{volume}{27}, \bibinfo{pages}{105--112}.
\bibitem[{Guerra and Centeno(2008)}]{GUERRA2008}
\bibinfo{author}{Guerra, M.}, \bibinfo{author}{Centeno, M.},
  \bibinfo{year}{2008}.
\newblock \bibinfo{title}{Optimal reinsurance policy: The adjustment
  coefficient and the expected utility criteria}.
\newblock \bibinfo{journal}{Insurance: Mathematics and Economics}
  \bibinfo{volume}{42}, \bibinfo{pages}{529 -- 539}.
\newblock \DOIprefix\doi{https://doi.org/10.1016/j.insmatheco.2007.02.008}.
\bibitem[{Hu et~al.(2015)Hu, Yang and Zhang}]{Hu2015}
\bibinfo{author}{Hu, X.}, \bibinfo{author}{Yang, H.}, \bibinfo{author}{Zhang,
  L.}, \bibinfo{year}{2015}.
\newblock \bibinfo{title}{Optimal retention for a stop-loss reinsurance with
  incomplete information}.
\newblock \bibinfo{journal}{Insurance: Mathematics and Economics}
  \bibinfo{volume}{65}, \bibinfo{pages}{15--21}.
\newblock \DOIprefix\doi{10.1016/j.insmatheco.2015.08.005}.
\bibitem[{Huberman et~al.(1983)Huberman, Mayers and Smith}]{Huberman1983}
\bibinfo{author}{Huberman, G.}, \bibinfo{author}{Mayers, D.},
  \bibinfo{author}{Smith, C.W.}, \bibinfo{year}{1983}.
\newblock \bibinfo{title}{Optimal insurance policy indemnity schedules}.
\newblock \bibinfo{journal}{The Bell Journal of Economics}
  \bibinfo{volume}{14}, \bibinfo{pages}{415--426}.
\newblock \URLprefix \url{http://www.jstor.org/stable/3003643}.
\bibitem[{Kaluszka(2001)}]{kaluszka2001}
\bibinfo{author}{Kaluszka, M.}, \bibinfo{year}{2001}.
\newblock \bibinfo{title}{Optimal reinsurance under mean-variance premium
  principles}.
\newblock \bibinfo{journal}{Insurance: Mathematics and Economics}
  \bibinfo{volume}{28}, \bibinfo{pages}{61--67}.
\bibitem[{Lo(2017)}]{lo2017unifying}
\bibinfo{author}{Lo, A.}, \bibinfo{year}{2017}.
\newblock \bibinfo{title}{A unifying approach to risk-measure-based optimal
  reinsurance problems with practical constraints}.
\newblock \bibinfo{journal}{Scandinavian Actuarial Journal}
  \bibinfo{volume}{2017}, \bibinfo{pages}{584--605}.
\bibitem[{Meng et~al.(2023)Meng, Wei and Zhou}]{MENG2023}
\bibinfo{author}{Meng, H.}, \bibinfo{author}{Wei, L.}, \bibinfo{author}{Zhou,
  M.}, \bibinfo{year}{2023}.
\newblock \bibinfo{title}{Multiple per-claim reinsurance based on maximizing
  the lundberg exponent}.
\newblock \bibinfo{journal}{Insurance: Mathematics and Economics}
  \bibinfo{volume}{112}, \bibinfo{pages}{33--47}.
\newblock \URLprefix
  \url{https://www.sciencedirect.com/science/article/pii/S0167668723000471},
  \DOIprefix\doi{https://doi.org/10.1016/j.insmatheco.2023.05.009}.
\bibitem[{Sun et~al.(2017)Sun, Weng and Zhang}]{sun2017optimal}
\bibinfo{author}{Sun, H.}, \bibinfo{author}{Weng, C.}, \bibinfo{author}{Zhang,
  Y.}, \bibinfo{year}{2017}.
\newblock \bibinfo{title}{Optimal multivariate quota-share reinsurance: A
  nonparametric mean-cvar framework}.
\newblock \bibinfo{journal}{Insurance: Mathematics and Economics}
  \bibinfo{volume}{72}, \bibinfo{pages}{197--214}.
\bibitem[{Tan et~al.(2020)Tan, Wei, Wei and Zhuang}]{TAN2020}
\bibinfo{author}{Tan, K.S.}, \bibinfo{author}{Wei, P.}, \bibinfo{author}{Wei,
  W.}, \bibinfo{author}{Zhuang, S.C.}, \bibinfo{year}{2020}.
\newblock \bibinfo{title}{Optimal dynamic reinsurance policies under a
  generalized denneberg’s absolute deviation principle}.
\newblock \bibinfo{journal}{European Journal of Operational Research}
  \bibinfo{volume}{282}, \bibinfo{pages}{345--362}.
\newblock \URLprefix
  \url{https://www.sciencedirect.com/science/article/pii/S0377221719307234},
  \DOIprefix\doi{https://doi.org/10.1016/j.ejor.2019.08.053}.
\bibitem[{Tan and Weng(2014)}]{tan2014empirical}
\bibinfo{author}{Tan, K.S.}, \bibinfo{author}{Weng, C.}, \bibinfo{year}{2014}.
\newblock \bibinfo{title}{Empirical approach for optimal reinsurance design}.
\newblock \bibinfo{journal}{North American Actuarial Journal}
  \bibinfo{volume}{18}, \bibinfo{pages}{315--342}.
\bibitem[{Wang and B{\o}lviken(2022)}]{wang2022much}
\bibinfo{author}{Wang, Y.}, \bibinfo{author}{B{\o}lviken, E.},
  \bibinfo{year}{2022}.
\newblock \bibinfo{title}{How much is optimal reinsurance degraded by error?}
\newblock \bibinfo{journal}{North American Actuarial Journal}
  \bibinfo{volume}{26}, \bibinfo{pages}{283--297}.
\bibitem[{Xu et~al.(2019)Xu, Zhou and Zhuang}]{Xu2019}
\bibinfo{author}{Xu, Z.Q.}, \bibinfo{author}{Zhou, X.Y.},
  \bibinfo{author}{Zhuang, S.C.}, \bibinfo{year}{2019}.
\newblock \bibinfo{title}{Optimal insurance under rank-dependent utility and
  incentive compatibility}.
\newblock \bibinfo{journal}{Mathematical Finance} \bibinfo{volume}{29},
  \bibinfo{pages}{659--692}.
\newblock \URLprefix
  \url{https://onlinelibrary.wiley.com/doi/abs/10.1111/mafi.12185},
  \DOIprefix\doi{https://doi.org/10.1111/mafi.12185},
  \href{http://arxiv.org/abs/https://onlinelibrary.wiley.com/doi/pdf/10.1111/mafi.12185}{{\tt
  arXiv:https://onlinelibrary.wiley.com/doi/pdf/10.1111/mafi.12185}}.
\bibitem[{Young(2006)}]{young2006premium}
\bibinfo{author}{Young, V.R.}, \bibinfo{year}{2006}.
\newblock \bibinfo{title}{Premium Principles}.
\newblock \bibinfo{publisher}{Encyclopedia of Actuarial Science}.
\bibitem[{Zheng and Cui(2014)}]{zheng2014optimal}
\bibinfo{author}{Zheng, Y.}, \bibinfo{author}{Cui, W.}, \bibinfo{year}{2014}.
\newblock \bibinfo{title}{Optimal reinsurance with premium constraint under
  distortion risk measures}.
\newblock \bibinfo{journal}{Insurance: Mathematics and Economics}
  \bibinfo{volume}{59}, \bibinfo{pages}{109--120}.
\bibitem[{Zhuang et~al.(2016)Zhuang, Weng, Tan and Assa}]{zhuang2016marginal}
\bibinfo{author}{Zhuang, S.C.}, \bibinfo{author}{Weng, C.},
  \bibinfo{author}{Tan, K.S.}, \bibinfo{author}{Assa, H.},
  \bibinfo{year}{2016}.
\newblock \bibinfo{title}{Marginal indemnification function formulation for
  optimal reinsurance}.
\newblock \bibinfo{journal}{Insurance: Mathematics and Economics}
  \bibinfo{volume}{67}, \bibinfo{pages}{65--76}.

\end{thebibliography}

	\appendix
\section{Proof of scale invariance}\label{sec:a}
It was claimed in Section \ref{sec:2.3} that the problem is invariant
with respect to $\xi=E(X)$.
To verify this note that
the expected profit when $I_\xi$  is the risk retained by the insurer is 
\begin{displaymath}
	{\cal G}_\xi(I_\xi)=\gamma\xi-\int_0^\infty K\{F_\xi(x)\}dI_\xi(x)
	=\gamma\xi-\int_0^\infty K\{F_1(x/\xi)\}d\{\xi I_1(x/\xi)\},
\end{displaymath}  
and when $x=\xi y$ is substituted in the integral
\begin{displaymath}
	{\cal G}_\xi(I_\xi)
	=\gamma\xi-\xi\int_0^\infty K\{F_1(y)\}dI_1(y)
	\hspace*{1cm}\mbox{so that}\hspace*{1cm}
	{\cal G}_\xi(I_\xi)
	={\cal G}_1(I_1)\xi.
\end{displaymath}
But the $1-\epsilon$ percentile of $X$ is 
$x_{\xi\epsilon}=\xi x_{1\epsilon}$, which means that the  Value at Risk
of the insurer is
\begin{displaymath}
	\mbox{VaR}_\xi(R_{I_\xi})=x_{\xi\epsilon}-I_\xi(x_{\xi\epsilon})=\xi x_{1\epsilon}-
	\xi I_1(x_{1\epsilon})= \mbox{VaR}_1(R_{I_1})\xi,
\end{displaymath}
and $\xi$ cancels in the ratio
${\cal C}_\xi(I_\xi)={\cal G}_\xi(I_\xi)/\mbox{VaR}_\xi(I_\xi)$.
The argument
when CVaR replaces VaR is the same. 
\section{Proof of~(\ref{e212})}\label{sec:b}
The loss $X$ in a portfolio of independent risks
has expectation and
standard  deviation $E(X)=\xi$ and
$\mbox{sd}(X)=c\sqrt{\xi}+o(\sqrt{\xi})$, and
the distribution function
and $1-\epsilon$ percentile are
\begin{equation}
	F(x)=\Phi\left(\frac{x-\xi}{c\sqrt{\xi}}\right)+o(1)
	\hspace*{1cm}\mbox{and}\hspace*{1cm}
	x_\epsilon=\xi+c\sqrt{\xi}\phi_\epsilon+o(\sqrt{\xi}),
	\label{a1}
\end{equation}
where $\Phi(x)$ is the standard normal integral and $\phi_\epsilon$ its 
$1-\epsilon$-percentile. Suppose everything up to $x_\epsilon$
has been reinsured though the contract
$I_{0x_\epsilon}$ for which~(\ref{e24}) (with $\beta=0$) and~(\ref{e27}) yield
\begin{equation}
	{\cal G}(I_{0x_\epsilon})=\gamma\xi-\int_0^{x_\epsilon}K\{F(x)\}dx
	=\gamma\xi-c\sqrt{\xi}\int_{-\sqrt{\xi}/c}^{\phi_\epsilon}K\{\Phi(y)\}dy
	+o(\sqrt{\xi}),
	\label{a2}
\end{equation}
after having inserted~(\ref{a1}) for
$F(x)$ and $x_\epsilon$ and
substituted$y=(x-\xi)/(c\sqrt{\xi})$ in the integral. These
operations leave the error term as stated in~(\ref{a2}) since 
$K(u)$ is bounded function.  
Hence,
when applying l'H$\hat{\mbox{o}}$pital's rule
\begin{displaymath}
	\frac{{\cal G}(I_{0x_\epsilon})}{\xi}
	=\gamma-\frac{c}{\sqrt{\xi}}\int_{-\sqrt{\xi}/c}^{\phi_\epsilon}K\{\Phi(y)\}dy+o(1)
	\longrightarrow \gamma-K(0)
	\hspace*{1cm}\mbox{as}\hspace*{1cm}\xi\rightarrow\infty,
\end{displaymath}
which is~(\ref{e212}).
\section{Calculations for Section \ref{sec:4}}\label{sec:c}
{\em Formula~(\ref{e52}).}
To show that ${\cal L}(I)$ in~(\ref{e51})
can be rewritten as in~(\ref{e52})        
first note that 
the  expected profit of the insurer
when using~(\ref{e24}) (with $\beta=0$) and~(\ref{e27})
becomes
\begin{displaymath}
	{\cal G}(I)=\gamma E(X)-\{\pi(I)-E\{I(X)\}
	=\gamma E(X)-\int_0^\infty K\{F(x)\}dI(x),
\end{displaymath}
so that
\begin{displaymath}
	{\cal L}(I)
	=\gamma E(X)-\int_0^\infty K\{F(x)\}dx-\mu\rho(I).
\end{displaymath}
The VaR risk measure for
$R_I(x)=x-I(x)$ is 
$\rho(I)=x_\epsilon-I(x_\epsilon)$
since $R_I(x)$ is non-decreasing, and with
$\cal I$ designating the indicator function 
\begin{displaymath}
	\rho(I)=x_\epsilon-\int_0^\infty {\cal I}_{x<x_\epsilon}dI(x),
\end{displaymath}
which yields~(\ref{e52}) and~(\ref{e53}) when the expression for $\rho(I)$
is inserted.
\\\\
{\em Formula~(\ref{e55}).}
The expression for ${\cal G}(I)$ is the same as above, but
$\rho(I)$ is different. Now 
\begin{displaymath}
	\rho(I)=E\{R_I(X)|X\geq x_\epsilon)=E\{X|X\geq x_\epsilon\}-E\{I(X)|X\geq x_\epsilon\},
\end{displaymath}
where
\begin{displaymath}
	E\{I(X)|X\geq x_\epsilon\}=\frac{1}{\epsilon}\int_{x_\epsilon}^\infty I(x)dF(x)
	=I(x_\epsilon)+
	\frac{1}{\epsilon}\int_{x_\epsilon}^\infty\{(1-F(x)\}dI(x)
\end{displaymath}
after integration by parts. Hence
\begin{displaymath}
	\rho(I)=E\{X|X\geq x_\epsilon\}-\int_0^\infty {\cal I}_{x<x_\epsilon}dI(x)
	-\frac{1}{\epsilon}\int_{x_\epsilon}^\infty\{(1-F(x)\}dI(x),
\end{displaymath}
which when inserted into~(\ref{e51}) with the expression for
${\cal G}(I)$ yields~(\ref{e55}) and~(\ref{e56}).

\end{document}